\documentstyle[epsfig]{elsart}

\begin{document}

\vspace*{1cm}
\begin{flushright}
 {\bf UCY-PHY-97/01}
\end{flushright}

\begin{frontmatter}
 
\vspace*{1cm}

\title{A versatile dielectron trigger for nucleon-nucleon and nucleus-nucleus 
collisions.}

\author{R.~Schicker\thanksref{corr}},
\author{M.~Axiotis},
\author{H.~Tsertos}
\address{University of Cyprus, Nicosia, Cyprus}

\begin{abstract}

A novel approach for a versatile first level dielectron trigger is presented. 
This trigger operates in the low multiplicity environment of nucleon-nucleon
reactions as well as in the high multiplicity situation of nucleus-nucleus
collisions. For optimal trigger performance, time of flight conditions for the 
two fastest particles of the event are combined with event multiplicity 
requirements. The dielectron trigger efficiency is given. The event reduction 
factor of such a trigger approach is studied for a low, a medium and a 
high multiplicity environment. The impact parameter dependence of the event
reduction is given. The timing properties of the trigger signal are 
described. The losses due to deadtime are specified. Finally, the first level 
trigger rate is reported.

\end{abstract}

\thanks[corr]{Corresponding author, e-mail "schicker@alpha2.ns.ucy.ac.cy" \\
Dept. Nat. Science, Univ. Cyprus, PO 537, 1678 Nicosia, Cyprus}

\end{frontmatter}

\newpage

\section{Introduction} 
\label{sec:intro}

Dielectrons are of particular interest in the study of nucleon-nucleon and 
nucleus-nucleus collisions since they do not undergo strong final state 
interactions. The measurement of dielectron observables thus reveals information 
which otherwise cannot be obtained by measuring hadronic degrees of 
freedom\cite{work1,work2}. Dielectrons are, however, produced with cross sections 
which are typically 4-5 orders of magnitude smaller than pion production cross 
sections. A large geometrical spectrometer acceptance and maximum beam intensity 
are therefore necessary in order to measure dielectron observables efficiently. 
Redundant lepton identification is mandatory for measuring the small dielectron
signal in the large hadronic background. In particular, the identification of 
electrons at the trigger level requires a sophisticated trigger scheme. 
The rate reduction required from the trigger system can only be achieved by a 
multi-level trigger architecture. 

\section{Trigger concepts}
\label{tconc}
\subsection{Conventional triggers}
\label{convt}

A first level trigger is usually designed as a reaction trigger. In low 
multiplicity environments such as in hadron induced reactions, a multiplicity 
condition of two can be used for the first level trigger\cite{DLSNIM,Ceres1}. This 
minimum condition of two represents the two partners of the pair. In high
multiplicity environments such as in heavy-ion induced collisions, different 
multiplicity conditions can be used to define a trigger. This trigger tags the 
events according to their charged particle multiplicity, and hence allows to 
accept or reject event classes of a given impact parameter range. A minimum 
multiplicity condition enables the measurement of the minimum bias multiplicity 
distribution. This minimum bias distribution is used to correct 
for the multiplicity trigger bias in order to study the impact parameter 
dependence of the dielectron yield\cite{DLSNb}.  
 
More restrictive trigger decisions can be made by using detector systems which 
combine a high electron detection efficiency with a high hadron rejection power
as is the case in Cherenkov detectors, for example. If the 
detector signal is prompt as in threshold gas Cherenkov counters, then this 
information can be used in the first level trigger\cite{DLSNIM}. If a higher 
hadron rejection power is needed, then Ring Imaging Cherenkov counters (RICH) 
can be used. However, the time scale of the ring pattern search does not allow 
the inclusion of RICH information in the first level trigger decision. Thus, 
RICH detector information is generally used in higher level 
triggers\cite{CeresCh,HadesCh}.

Electron identification on the trigger level can alternatively be achieved by 
using time of flight (TOF) or calorimetric information. The complexity of 
electron identification in a large hadronic background, however, restricts TOF 
or calorimeter information in general to higher level trigger 
decisions\cite{Hadesprop,taps}.

\subsection{The electron time of flight trigger}
\label{toft}

The new trigger concept presented in this paper allows a much improved first 
level trigger performance by integrating preliminary electron identification into
the first level trigger. This preliminary electron identification is based on 
isochronous TOF information, and is therefore complementary to the information 
from prompt Cherenkov counters. Thus, this novel approach combines a conventional
first level reaction trigger with aspects of traditional higher level electron 
triggers. Moreover, this new first level trigger approach operates in low as 
well as in medium to high multiplicity environments.

This novel electron TOF trigger is of practical interest for dielectron 
spectrometers of any magnetic field configuration. A highly segmented TOF system
is, however, a prerequisite for spectrometers of large geometrical acceptance. 
This TOF segmentation is required for optimizing electron isochronicity within 
the solid angle covered by the spectrometer.   
Electron isochronicity is achieved by adding delays to the signal of each TOF
segment. These delays compensate for the variations in trajectory length 
within the acceptance of the spectrometer. The delayed signals are, however,
prompt on the time scale of the first level trigger. 

\subsection{Electron time of flight trigger simulations}
\label{tofts}

In order to study the performance of such a trigger approach, simulations were 
carried out by taking the dielectron spectrometer HADES at GSI as a specific 
example. The HADES experiment combines a large geometrical acceptance with a 
highly segmented TOF array\cite{hades1}. The performance of such a trigger 
architecture is studied for a low, a medium and a high multiplicity environment.
In particular, the performance of this new trigger approach is compared to the 
performance of the corresponding first level trigger which does not identify 
electron candidates.

HADES is designed to operate at heavy-ion and proton beam intensities of 
10$^{8}$ s$^{-1}$. A 1\% interaction target results in 10$^{6}$ s$^{-1}$ 
minimum bias events. The maximum available pion beam intensity, on the other 
hand, reaches values of about 2$\times$10$^{7}$ per spill at a momentum of 
1 GeV/c. The proton beam is therefore more demanding on the first level trigger 
as compared to the pion beam. Thus, simulations have been carried out only for 
heavy-ion and proton induced reactions. The average charged particle multiplicity 
is higher in proton-nucleus than in proton-proton collisions, and hence heavy-ion
targets are more demanding on the trigger system as compared to proton targets. 
Simulation results are therefore shown below for the systems p+Ni at 2 AGeV, 
Ne+Ne at 2 AGeV and Au+Au at 1 AGeV as examples of a low, a medium and a high 
multiplicity environment, respectively. Studies of the high multiplicity system 
Ni+Ni confirm the conclusions drawn from the above three systems\cite{axio}. 

This paper is organized as follows: Section \ref{sec:ftrig} gives a summary of 
first level trigger requirements. In Section \ref{sec:pdat}, the 
simulation of the first level trigger data is described. Section \ref{sec:adat} 
introduces the analysis of the simulation data. Section \ref{sec:fpni} reports 
on the first level trigger in the low multiplicity reaction p+Ni. 
Section \ref{sec:fnene} gives details of the first level trigger performance 
in the medium  multiplicity system Ne+Ne. In Section \ref{sec:fauau}, the 
performance characteristics of the first level trigger in the high multiplicity 
system Au+Au is presented. 

\section{First level trigger requirements}
\label{sec:ftrig}

\subsection{Hadron induced reactions}
\label{ssec:fhadr}

A reaction trigger for hadron induced collisions can be defined by requiring a 
multiplicity condition of two or larger. This minimal multiplicity value of two 
corresponds to the two charged leptons of the pair. To illustrate this point, 
we mention here the p($\pi^{-}$,e$^{+}$e$^{-}$)n reaction which is tagged by a 
multiplicity condition of two. This reaction attracts considerable interest for 
the investigation of the time-like nucleon form factor below the threshold 
accessible in nucleon-antinucleon annihilation. If the reaction channel to
be measured contains additional charged tracks, then the imposed multiplicity
condition can be correspondingly increased.  Thus, a versatile first level 
trigger for hadron induced reactions is characterized by a multiplicity 
condition of two or larger. 

In the HADES spectrometer, the multiplicity condition for the first level 
trigger can be derived from the highly segmented TOF system. In combination with 
a timing signal of an additional start detector, the TOF trigger signal is used 
to establish gates for ADCs and to define START/STOP signals for TDCs. Hence the
delay of the TOF trigger signal with respect to the time of reaction as well as 
the time jitter are of particular interest. The time jitter of the first level 
trigger signal arises from different sources. First, velocity variations of the 
particles defining the trigger transition induce particle time of flight 
variations. Second, trajectory length variations over the acceptance of the 
spectrometer add to the time of flight variations. Third, different signal 
propagation times in the TOF elements depending on location of the hit point 
add varying delays to the TOF signals. 

A multiplicity condition of two is triggered by the two fastest particles of the 
event. For events containing dielectrons, the two fastest particles are the two 
electrons ($\gamma \geq 20$) of the pair. The contribution to the time jitter 
due to particle velocity variations mentioned above therefore vanishes. Thus, 
for a multiplicity condition of two, the trigger time jitter of dielectron 
events is considerably reduced as compared to events without dielectrons. This 
time correlation allows to define a timing window $\Delta$T$_{0}$ for the trigger 
signal derived from the two fastest particles of the reaction. Events containing
a dielectron will meet this condition, but most of the events without dielectrons
will not. The trigger time jitter of dielectron events therefore contains only 
the contributions of trajectory lengths variations and of different signal 
propagation delays as described above. In a highly segmented TOF system, however,
trajectory length variations within one TOF paddle are very much reduced as 
compared to the variations within the full spectrometer acceptance. The 
differences in the mean TOF value from paddle to paddle can, however, be 
corrected by cable delays. Thus, the mean electron TOF values (including cable 
delays) of all TOF paddles are isochronous. The remaining trigger time jitter is 
dominated by the varying signal propagation delays depending on the hit location 
within the TOF paddle. In a paddle geometry with readout on both sides, this 
jitter contribution can, however, be minimized by a mean timer circuit. 

The cable delays of the individual TOF paddles and the mean timer improve the 
trigger time correlation of dielectron events considerably. The timing window 
$\Delta$T$_{0}$ described above can therefore be narrowed, and less events 
without dielectrons will initiate a trigger. The deadtime of the first level 
trigger is  therefore reduced considerably. Such a timing window $\Delta$T$_{0}$
necessitates, however, an independent measurement of the reaction time by 
another detector system. This additional detector system is the start detector 
mentioned above.

\subsection{Heavy-ion induced reactions}
\label{ssec:fheavy}

A first level trigger in heavy-ion induced reactions tags the events according 
to their charged particle multiplicity. The correlation between impact parameter
and charged particle multiplicity enables the selection of impact parameter 
ranges by defining corresponding multiplicity conditions. This multiplicity can 
be derived from a highly segmented detector system. In the case of the HADES 
spectrometer, the highly segmented TOF array is used. 

The six fold azimuthal 
segmentation of the HADES setup allows for two possible definitions of a 
multiplicity trigger. A lower limit on multiplicity in each of the six sectors 
defines a sector multiplicity condition. Alternatively, a lower limit on the 
multiplicity summed over the six sectors defines a total multiplicity condition. 
For central heavy-ion reactions, these two different multiplicity conditions 
are expected to be equivalent due to the azimuthal symmetry of the reaction
zone. For semi-central or peripheral reactions, however, these two conditions 
are not equivalent due to azimuthal anisotropies of the reaction.

The minimal sector multiplicity for defining  a trigger is equal to one.
Sector multiplicity conditions can therefore be used for systems which have a 
small probability to have no tracks in one or more of the azimuthal segments. 
We define a high multiplicity environment by the property that a sector 
multiplicity condition of one results in an efficiency loss of less than 10\% 
for central events. These central events exhibit azimuthal symmetries to a 
high degree. The multiplicity distribution of one azimuthal segment can 
therefore be approximated by a Binomial distribution B(n$_S$,n$_T$,p).
Here, n$_S$ and n$_T$ are the multiplicity in one segment and the total
multiplicity, respectively. The parameter p represents the probability 
that a single track is located in the segment under consideration. 
In absence of azimuthal correlations and in the case of the HADES spectrometer,
the parameter p is equal to 1/6.

The probability to have $n_S$ tracks in a sector is given by 

\begin{displaymath}
P(n_S) = \frac{n_T!}{(n_T-n_S)!n_S!} \cdot \frac{5^{n_T-n_S}}{6^{n_T}}
\end{displaymath}

If the minimal sector multiplicity condition of one is used, then efficiency 
losses result for events which have multiplicity zero in one or more of the 
azimuthal segments. In leading order, this probability $W_{Loss}$ is given by 
the number of sectors multiplied by the probability of no tracks in a sector.

\begin{displaymath}
W_{Loss} = 6 \cdot \left. P(n_S) \right|_{n_S=0} = 6\cdot (\frac{5}{6})^{n_T} 
\end{displaymath}

High multiplicity events are defined by efficiency losses smaller than 10\%, 
hence $n_T$ of these events is determined by
  
\begin{displaymath}
6\cdot (\frac{5}{6})^{n_T} \leq 0.1
\end{displaymath}

This inequality holds for $n_T \geq \frac{ln\,60}{ln\,60-ln\,50 }$, i.e. for 
$n_T \geq 23$.

Taking into account the HADES solid angle of about 50\%, events with maximum 
total multiplicity of about $M_{Tot}^{max} \geq 50$ are defined as high 
multiplicity events. Accordingly, events with $M_{Tot}^{max} \leq 50$
are defined as medium multiplicity systems. In medium multiplicity systems,
first level trigger conditions can only be applied to the total multiplicity.

Comprehensive dielectron spectroscopy of heavy-ion collisions necessitates the 
measurement of the dielectron signal as a function of impact parameter of the 
reaction. With the first level trigger approach presented here, data of central,
semi-central and peripheral events can be taken by redefining the multiplicity
conditions and by appropriately downscaling the resulting trigger rate. Thus, 
non-central events can either be downscaled and be registered simultaneously
with central events or can be recorded in dedicated data taking periods.
The major interest is, however, focused on central heavy-ion reactions due 
to the large densities reached in these collisions. 
 
Each signal of the first level trigger has an associated deadtime due to 
frontend readout of detectors. This deadtime is about 10 $\mu$sec. 
If the multiplicity condition is set low, then the trigger rate will be 
increased with a corresponding increase in probability that central events 
fall into the deadtime window. If, on the other hand, the multiplicity 
condition is set high, then central events start to get rejected due 
to insufficient multiplicity. For a given deadtime, there exists 
therefore a multiplicity condition which optimizes the number of central 
events which are passed onto the next trigger stage.

Due to the statistical occurrence of heavy-ion reactions, there is a finite
probability that two or more events occur very close in time. Hence, TOF paddles 
can carry simultaneously signals of different events. The combined signals of 
the different events may satisfy the trigger condition whereas none of the 
individual events would be able to do so. This overlap probability depends 
strongly on the reaction rate and, thus, on beam intensity. For all of the 
results shown, a beam rate of 10$^{8}$ s$^{-1}$ and a minimum bias event
rate of 10$^{6}$ s$^{-1}$ is assumed.

The performance of the first level trigger in heavy-ion collisions depends weakly 
on the duration of the TOF paddle signals\cite{UCY1,UCY2}. If the TOF signal 
length is short, then the signals of the fastest particles have disappeared
before the signals of the slower particles have started. The trigger system 
will therefore see an apparently reduced event multiplicity with increased 
multiplicity fluctuations. TOF signals which are long, on the other hand, 
result in an increased number of triggers from events overlapping in time. 
In this report, all the results for heavy-ion induced reactions have been 
derived with a TOF signal duration of 15 nsec. 

\section{First level trigger data simulation}
\label{sec:pdat}

For studying the behavior of the first level trigger, the full HADES geometry
was implemented into the GEANT package\cite{Heike}. A realistic field map of 
the toroidal magnetic field is used for tracking of the charged particles.

The collisions of the three systems studied, p+Ni, Ne+Ne and  Au+Au are modeled
by a transport equation of the Boltzmann-Uehling-Uhlenbeck (BUU) type.
The dynamical evolution of the collisions is determined by calculating
the phase space evolution for nucleons, Delta and N$^{*}$ resonances.
With this code, good agreement is found between data and model predictions
for nucleon, pion, kaon and dilepton distributions in proton and heavy-ion
induced reactions in the energy range 1-2 AGeV\cite{Wolf}. Since the charged 
particle multiplicity is mainly due to protons and pions, only these two 
particle species are tracked for the first level trigger simulations.

For the simulations of the first level trigger in these three  systems,
BUU events of different impact parameters are used. In these systems, the upper 
limit of the impact parameter range is defined by the geometrical cross section.
In each of the three systems studied, six to eight discrete equidistant impact 
parameters represent the full range from zero to maximum value. Each tracked 
particle of an event is followed through the complete HADES geometry. A 
trajectory entering a TOF paddle volume defines a TOF hit. The information of 
this hit contains the TOF paddle number and a TOF time value. This time value
represents the sum of the particles time of flight from target to the paddle 
hit point plus the signal propagation delay within the TOF paddle. In the left 
or right (LOR) timing, this signal delay is determined by the shorter of the 
two propagation times of the signal to either end of the paddle. In the MEAN 
timing, the signal propagation delay is independent of particle hit point and 
is equal to the propagation delay through half the length of the paddle. In the 
calculation of the signal propagation delay, signal time jitter or finite 
resolution effects are neglected. The exact TOF time value is subsequently 
converted into an integer format with a discretization accuracy of 10 psec and 
written into a CERN CWN-tupel. At this stage, each discrete impact parameter 
is simulated separately. Hence, separate tupel files exist for each of the 
discrete impact parameters simulated.

\section{First level trigger data analysis}
\label{sec:adat}

In the analysis of the first level trigger, a stream of events is 
generated by assigning a random time interval t$_{\Delta}$ 
and a random impact parameter b to each event. 

The time interval t$_{\Delta}$ is chosen according to the probability
$p(t_{\Delta}) = \lambda \cdot \e^{-\lambda\cdot t_{\Delta}}$. This 
probability distribution represents the distribution of time intervals
between two consecutive events, with $\lambda$ being equal to the
inverse of the average time between two consecutive events. The 
absolute time of an event is determined by adding t$_{\Delta}$ to the 
absolute time of the previous event.
                               
The impact parameter b of an event is taken from a distribution which is 
zero at b\,=\,0\,fm and linearly increasing up to b\,=\,b$_{max}$. Here, 
b$_{max}$ is defined by the geometrical cross section as explained above. 
An event is subsequently read from the tupel file with the closest nearby 
simulated impact parameter. 

The analysis of the first level trigger loops over all TOF paddles in time steps
of 1 nsec. In each loop, the sum of the TOF paddles carrying a signal at that 
particular moment is calculated. If the given trigger multiplicity requirement 
is met in a step, but not in the previous one, then a trigger transition is 
induced. The exact timing of the trigger transition is determined from the 
information of the TOF time values to within the discretization accuracy of 
10 psec as described above. This trigger transition has to arrive within a 
trigger time window $\Delta$T$_{0}$ as described in Section \ref{ssec:fhadr}.
In the prompt trigger described below, a trigger
transition generates a trigger. In the delayed trigger described below, a 
trigger transition initiates a time window of 20 nsec during which the event 
has to satisfy an additional multiplicity requirement. If this additional 
condition is met, then a trigger is generated. Here, prompt and delayed 
trigger refer to the one and two step trigger approach as described in 
Section \ref{ssec:fmt}. Each trigger starts a deadtime window during which 
no other triggers are accepted. 

\section{First level trigger in p+Ni collisions}
\label{sec:fpni}

\subsection{Minimum total multiplicity}
\label{sec:fpnimt}

Due to the low multiplicity of the p+Ni system, a large fraction of reactions 
has no tracks in one or more of the azimuthal sectors. A condition on minimal 
total multiplicity M$_{L} \geq 2$ is therefore used in order to define the 
trigger. Thus, the two fastest particles of an event define the trigger.   

\subsection{Trigger timing}
\label{sec:fpnitt}

The dashed line in Fig. \ref{fig:fig1} shows the LOR trigger timing of p+Ni 
events summed over impact parameters. The time zero is the time of reaction. This
timing distribution has a considerable tail at high $\Delta$T$_{0}$ values. This 
tail results from events in which the second particle of the trigger condition 
originates from target fragmentation and is very slow. Due to these events, the 
signal length of the TOF paddles is taken to be 50 nsec in the p+Ni system. A 
TOF signal value less than 50 nsec would, in addition to the geometrical 
spectrometer acceptance, introduce an additional trigger efficiency factor. 
The systematic errors of cross section measurements are therefore
minimal for a TOF signal length of 50 nsec or larger.

\subsection{Trigger timing of events containing dielectrons} 
\label{sec:fpnitte}

The width of the trigger timing shown in Fig. \ref{fig:fig1} by the dashed line 
results from three different sources as explained in Section \ref{ssec:fhadr}. 
The trigger timing of dielectron events does not contain the jitter 
contribution due to particle velocity variations. Thus, as shown in 
Fig. \ref{fig:fig1} by the solid line, the trigger time jitter is reduced for 
dielectron events. The FWHM  of this dielectron  timing distribution is 
about 3 nsec. In Fig. \ref{fig:fig1}, the 
events without dielectrons are downscaled with respect to the dielectron events. 
The relative intensity of these two classes of events is therefore arbitrary.

The timing correlation of dielectron events allows to reject trigger transitions  
which do not arrive within a time window  $\Delta$T$_{0}$ following the reaction.
This time window $\Delta$T$_{0}$ is defined such that about 98\% of the 
dielectron events are contained within the window. The vertical dotted line in 
Fig. \ref{fig:fig1} manifests such a trigger time window. Incidentally, the 
boundary of this time window coincides approximately with the peak intensity of 
event triggers which do not contain dielectrons. Modifications of the dielectron
trigger timing studied below will therefore always be compared to the timing of 
the peak intensity of no-pair events.

As discussed in Section \ref{ssec:fhadr}, the dielectron trigger timing shown in 
Fig. \ref{fig:fig1} by the solid line results from trajectory length variations 
over the polar angle of the spectrometer and from variations in signal 
propagation delays in the TOF paddles. Fig. \ref{fig:fig2} displays these 
different trigger jitter contributions as a function of the paddle number in 
both the LOR and the MEAN timing approach. Low paddle numbers represent small 
polar angles, whereas high paddle numbers correspond to large polar angles. 
Shown in Fig. \ref{fig:fig2} is the TOF time information of single electrons
as a function of paddle number. The diamond symbols represent the values of 
TOF alone, without time delays due to signal propagation in the paddles.
The structure seen results from the chosen geometrical partitioning of the 
full polar angular range into two segments. The vertical error bars indicate 
the width of the TOF distribution of trajectories hitting the paddle.  
The data shown by the solid line represent the MEAN timing values. Here, the 
timing is defined by the TOF values plus the time delay introduced by the MEAN 
timer. The difference between these data points and the diamond symbols is 
therefore proportional to the geometrical length of the TOF paddles.
In these simulations, it is assumed that the MEAN timer circuit does
not introduce any time jitter. The error bars of the diamond and the 
solid line data are therefore equal. 
The dashed line in Fig. \ref{fig:fig2} depicts the LOR timing. If the particle 
hit distribution is uniform along the length of the TOF paddles, then the LOR 
timing should on average contain half the propagation delay as compared to the 
MEAN timing. This is indeed seen in Fig. \ref{fig:fig2} where the dashed line 
data of the LOR timing are located about midway between the TOF only and the 
MEAN time data. In the LOR timing, the signal propagation delay in the TOF 
paddles is position dependent. This position dependency broadens the time 
distribution of particles hitting a paddle. The error bars of the LOR timing 
are therefore considerably increased with respect to the MEAN timing error bars. 

The width of the trigger timing of dielectron events shown in Fig.
\ref{fig:fig1} by the solid line results from folding the differential 
timing distribution of Fig. \ref{fig:fig2} with the polar angular distribution
of the single electrons. The width of the dielectron trigger timing of 
Fig. \ref{fig:fig1} can be reduced by eliminating the polar angle dependence 
of the single electron timing shown in Fig. \ref{fig:fig2}. This can be 
achieved by introducing an individual cable delay for each paddle. This delay 
ranges from zero for the largest paddle numbers to about 5 nsec for the smallest
paddle numbers. We define the corrected LOR timing as the time resulting from 
the LOR timing corrected for the polar angle dependence by cable delays. 
Similarly, the corrected MEAN timing is the time from the MEAN timing 
corrected for the polar angle dependence by cable delays.
  
The upper part of Fig. \ref{fig:fig3} shows the trigger time in the corrected 
LOR timing approach. These data correspond to the data of Fig. \ref{fig:fig1},
but with the improved trigger timing as explained above. 
The width of dielectron triggers is now reduced from the 3 nsec shown in
Fig. \ref{fig:fig1} to about 2.5 nsec. The peak intensity of the triggers from 
events without dielectrons is shown by the vertical dot-dashed line. 
Thus, this vertical dot-dashed line represents the trigger time window 
$\Delta$T$_{0}$ if no isochronous timing corrections are applied. 
As can be seen in this figure, the corrected LOR timing reduces the first level
trigger rate due to the improved dielectron trigger time window $\Delta$T$_{0}$.

The lower part of Fig. \ref{fig:fig3} shows the trigger time in the corrected 
MEAN timing approach. The width of dielectron triggers is now reduced from the 
3 nsec shown in Fig. \ref{fig:fig1} to about 0.5 nsec. As in the upper part of 
Fig. \ref{fig:fig3}, the vertical dot-dashed and dotted lines indicate the 
trigger time window $\Delta$T$_{0}$ without timing corrections and trigger time 
window $\Delta$T$_{0}$ for the corrected MEAN timing, respectively. The further 
improved trigger time window results in a considerable rate reduction as 
compared to the rate resulting from the time window $\Delta$T$_{0}$ shown in 
Fig. \ref{fig:fig1}.

Each trigger has an associated deadtime T$_{0}$ due to frontend readout of 
electronic channels. An event trigger occuring during the deadtime window of 
the previous trigger will not initiate readout and the information of the event 
is lost. Deadtime losses are quantified by introducing a factor R$_{DT}$.
Thus, the first level trigger rate is given by the reaction rate multiplied 
by the product of trigger efficiency times R$_{DT}$.

As explained above, a start detector is needed for an independent measurement of 
the time of reaction. This independent measurement introduces an additional time
jitter in the determination of the trigger timing. In the simulations presented 
here, the start detector time jitter has been taken into account by adding a 
Gaussian distributed uncertainty to the LOR and MEAN trigger timing. Table 
\ref{tab:tab1} lists trigger rates derived from the corrected MEAN timing for 
deadtimes T$_{0}$ = 0,6 and 10 $\mu$sec and for start detector resolutions of 
$\sigma_{Start}$ = 0.0, 0.1, 0.2 and 0.4 nsec. Here, events with minimum total
multiplicity M$_{L} \geq$ 2 generate a trigger. The first row shows the rates 
with a wide open trigger time window $\Delta$T$_{0}$= 60 nsec. As shown in
Fig. \ref{fig:fig1}, a condition $\Delta$T$_{0}$= 60 nsec is equivalent to 
no trigger window condition. The following rows display, for each value of start
detector resolution, the trigger rates and R$_{DT}$ values for trigger time 
windows $\Delta$T$_{0}$ which cut the dielectron events at the 98\% level (see 
Fig. \ref{fig:fig3}). For each deadtime T$_{0}$, the rate is shown in the left 
column in units of 10$^{5}$ s$^{-1}$. The R$_{DT}$ value is displayed in the 
corresponding column on the right. Without the trigger time window condition 
$\Delta$T$_{0}$, the primary rate of 4.8$\times$10$^{5}$ results in an R$_{DT}$
value of 0.17 for the expected deadtime T$_{0}$ = 10 $\mu$sec
Here, the primary rate is defined as the rate at zero deadtime. 
This R$_{DT}$ value of 0.17 indicates a first level trigger system deadtime of 
83\%. If the correct trigger window conditions $\Delta$T$_{0}$ are used for the 
different start detector time resolutions, the primary rate of 
4.8$\times$10$^{5}$ is reduced by a factor of about 20. For the expected values 
of $T_{0} =$ 10$\,\mu$sec and $\sigma_{Start}=$0.2 nsec, for example, the 
R$_{DT}$ value is equal to 0.82. Such a R$_{DT}$ value implies a system 
deadtime of 18\%. The system deadtime reduction from 83\% to 18\% represents 
an increase of the number of dielectron events passed to the next trigger stage 
by a factor of about five.
Thus, the new trigger approach presented here of integrating electron 
identification into the first level trigger reduces the beam time requirement 
for the system p+Ni by a factor of about five. 

\section{First level trigger in Ne+Ne collisions}
\label{sec:fnene}

\subsection{Trigger timing}

Fig. \ref{fig:fig4} shows the timing of the trigger for Ne+Ne events of 
different impact parameters. Here, the trigger time is defined by the condition 
of minimum total multiplicity M$_{L} \geq 2$. The time jitter contribution of 
the start detector is taken into account by adding a Gaussian distributed 
uncertainty of $\sigma_{Start} =$ .2 nsec to the trigger time. In the top part 
of Fig. \ref{fig:fig4}, the trigger time derived from the corrected LOR timing 
is displayed. In the lower part, the trigger time from the corrected MEAN timing
is shown. The trigger window condition $\Delta$T$_{0}$ described in Section 
\ref{ssec:fhadr} is indicated in the upper and lower part by the vertical dotted
lines. Such a trigger window condition $\Delta$T$_{0}$ reduces the event rate 
by less than 10\% in the LOR timing. In the MEAN timing, however, the event 
rate reduction achieved by the window condition $\Delta$T$_{0}$ is about 40\%.

\subsection{Minimum total multiplicity}

As described above, the multiplicity condition M$_{L} \geq 2$ together
with the trigger window condition $\Delta$T$_{0}$ does not reduce the rate
by the required factor of ten. Thus, the multiplicity condition M$_{L}$ 
needs to be increased in order to get the necessary rate reduction. 
Fig. \ref{fig:fig5} shows the trigger efficiency for Ne+Ne events as a 
function  of the total multiplicity condition M$_{L}$. Here, the trigger
window condition $\Delta$T$_{0}$ has been set wide open. The data shown 
in Fig. \ref{fig:fig5} represent therefore the trigger efficiency due to
multiplicity selection only. Fig. \ref{fig:fig5}  displays the data 
points for events with impact parameters of 1,2 and 3 fm, respectively.
For all the impact parameters shown in Fig. \ref{fig:fig5}, the efficiency 
exhibits a plateau of nearly 100\% at low total multiplicity values but drops 
steeply at large values. 

\subsection{Trigger rate}

Table \ref{tab:tab2} lists the trigger rates for the Ne+Ne system for deadtimes 
of T$_{0}$ = 0,6 and 10 $\mu$sec and for minimum total multiplicity conditions 
M$_{L}$ =2,3,4,5 and 6. As shown in Fig. \ref{fig:fig5}, multiplicity conditions 
M$_{L} > 6$ would reduce the efficiency for central events and are therefore not
considered here. For each deadtime T$_{0}$ in Table \ref{tab:tab2}, the rate is 
presented on the left in units of 10$^{5}$ s$^{-1}$ and the corresponding 
R$_{DT}$ value on the right. The first two rows of Table \ref{tab:tab2} show the 
rates resulting from a multiplicity condition M$_{L} \geq$2. The first row 
displays the rates if a wide open trigger window $\Delta$T$_{0}$ is applied, 
i.e. no trigger window condition. The second row shows the rates if the correct 
window condition $\Delta$T$_{0}$ is used as described in Section 
\ref{ssec:fhadr}. The R$_{DT}$ values of .21 and .31 indicate system deadtimes 
of about 80\% and 70\%, respectively. The following rows show the data for total
multiplicity conditions M$_{L} >$ 2. For these data, the event timing is 
determined from the two fastest particle of the event and an independent 
multiplicity condition M$_{L}$ is required. For the expected deadtime 
T$_{0}=10\,\mu$sec, for example, the multiplicity condition M$_{L} \geq 6$ 
results in system deadtimes of 20\% and 63\% with and without the trigger 
window condition $\Delta$T$_{0}$. These values represent an increase of 
the number of dielectron events passed to the next trigger stage by a factor of 
about two when applying the trigger window condition $\Delta$T$_{0}$ derived 
from the corrected MEAN timing.
Thus, the new trigger approach of integrating electron 
identification into the first level trigger reduces the beam time requirement 
for the system Ne+Ne by a factor of about two. 

\section{First level trigger in Au+Au collisions}
\label{sec:fauau}

Fig. \ref{fig:fig6} shows the timing of the trigger for Au+Au events of 
different impact parameters. Here, the trigger time is defined by the 
condition of minimum total multiplicity M$_{L} \geq 2$. The time jitter 
contribution of the start detector is taken into account by adding a Gaussian 
distributed uncertainty of $\sigma_{Start} =$ .2 nsec to the trigger time. 
The top and bottom part of Fig. \ref{fig:fig6} display the trigger time 
derived from the corrected LOR and MEAN timing, respectively. The trigger 
window condition $\Delta$T$_{0}$ described in Section \ref{ssec:fhadr} is 
indicated in the upper and lower part by the vertical dotted lines. Due to 
abundantly produced pions, such a trigger window condition $\Delta$T$_{0}$ 
has very little effect in the high multiplicity system Au+Au. In this system,
a first level trigger has to be based on multiplicity selection only. 

\subsection{Sector multiplicity condition}

In the high multiplicity system Au+Au, a multiplicity condition in 
each azimuthal sector will tag central events. Fig. \ref{fig:fig7} shows the 
tagging efficiency for this system as a function of the imposed sector 
multiplicity condition M$_{S}$. This condition implies that the charged 
particle multiplicity is greater or equal to M$_{S}$ in each of the six 
azimuthal sectors. Shown are the data points for events with impact parameters 
of 1,3 and 5 fm, respectively. For all the impact parameters shown in Fig. 
\ref{fig:fig7}, the efficiency exhibits a plateau of nearly 100\% at low 
sector multiplicity values but drops steeply at large multiplicity values. 
The optimal choice for the sector multiplicity condition is a value as large 
as possible but still within the plateau of the impact parameter b\,=\,1\,fm. 
Hence, the sector multiplicity condition is set to M$_{S}\geq\,8$ for the 
calculations shown below.

\subsection{Trigger timing}
\label{ssec:ftauau}

Events satisfying the sector multiplicity condition M$_{S}$ generate a trigger 
transition. The delay of this signal relative to the time of reaction and the 
signal jitter are of interest. Fig. \ref{fig:fig8} shows the corrected MEAN time
of the trigger transition for events with different impact parameters. Here, the 
two fastest particle in each azimuthal sector define the sector timing. The 
trigger time is then defined by the one of the six sector timings which arrives 
last. Central events are represented by the solid line. The FWHM of their time 
distribution amounts to about 1 nsec. The FWHM value for events with impact 
parameters of 3 and 5 fm are about 1.2 and 1.5 nsec, respectively. Semi-central 
events meet the condition M$_{L} \geq 2$ for the sector timing only with the 
help of slower moving particles. Hence, for increasing impact parameters, a 
shift of the centroid to larger time values as well as a broadening of the 
distribution is seen. The time zero in Fig. \ref{fig:fig8} is the 
time of reaction.

\subsection{Total multiplicity}

The sector multiplicity requirement defines a condition on minimum particle 
multiplicity in each of the six azimuthal sectors. A condition on minimum total 
multiplicity might further reduce non-central events while at the same time 
accepting all central events. At the moment of the trigger transition, the 
total multiplicity for central events is still building up, whereas it is 
almost exhausted for non-central events. Fig. \ref{fig:fig9} displays the 
maximum total multiplicity reached during a time window of 20 nsec following 
the trigger transition. Central and non-central events which satisfy the sector 
multiplicity condition M$_{S} \geq 8$ behave quite differently in Fig. 
\ref{fig:fig9}. A condition on minimum total multiplicity will therefore 
further reduce non-central events.

\subsection{Timing total multiplicity}

Fig. \ref{fig:fig10} shows the time at which the maximum event multiplicity is 
reached for events with impact parameters b=1,3 and 5 fm. Here, the time zero 
is the time of the trigger transition defined by the sector multiplicity 
condition M$_{S}$. For central events, the maximum total multiplicity 
develops between 5 and 15 nsec following the trigger transition. 
Non-central events with impact parameters b\,=\,5\,fm develop their
maximum total multiplicity during a time span of about 10 nsec following
the trigger transition. Hence a time window of 15-20 nsec duration
following the trigger transition seems adequate to test for the maximum 
total event multiplicity.

\subsection{Trigger efficiency}

Each trigger transition followed by a total multiplicity larger than a required 
minimum value M$_{T}$ generates a trigger. Each trigger has an associated 
deadtime due to frontend readout of electronic channels. An event trigger 
occuring during the deadtime window of the previous trigger will not initiate 
readout and the information of the event is lost. Fig. \ref{fig:fig11} shows 
the values of Eff$_{LV1} \times$ R$_{DT}$ for central events for M$_{S} \geq 8$ 
as a function of the required total multiplicity M$_{T}$. 
Here, Eff$_{LV1}$ denotes the first level trigger efficiency, i.e.,
the efficiency for zero deadtime. R$_{DT}$ is a reduction factor which
contains the effects of the trigger deadtime. 
For the deadtime of T$_{0} =$ 10 $\mu$sec, the Eff$_{LV1} \times$ R$_{DT}$ value
has a maximum at an M$_{T}$ value of about 110. At lower M$_{T}$ 
values, semi-central events do not get suppressed efficiently. Thus, 
the increased rate of semi-central events leads to a decrease of the  
Eff$_{LV1} \times$ R$_{DT}$ values of central events. For high
M$_{T}$ values, Eff$_{LV1} \times$ R$_{DT}$ decreases since central
events start to get rejected due to insufficient multiplicity.
The data for the deadtime of T$_{0} =$ 6 $\mu$sec are shown for comparison.   

\subsection{Trigger rate}

Table \ref{tab:tab3} lists the trigger rates in units of 10$^{5}$ s$^{-1}$ for 
the different total multiplicity thresholds M$_{T}$ and for deadtimes of T$_{0}$
= 0,6 and 10 $\mu$sec, respectively. A first level trigger rate of 1.$\times$
10$^{5}$ s$^{-1}$ is the input design value for the next trigger stage. Each 
value of T$_{0}$ in Table \ref{tab:tab3} corresponds to two columns. The left 
column shows the rate of trigger transitions resulting from the sector 
multiplicity condition M$_{S} \geq 8$. The right column displays the rate 
if additionally the total multiplicity condition M$_{T}$ is required. 

Table \ref{tab:tab4} lists the partial trigger rates from events of different 
impact parameters. For the chosen multiplicity requirements, triggers from
events with impact parameters b$\geq$5\,fm are negligible and therefore not 
listed in Table \ref{tab:tab4}. Both impact parameters b\,=\,1 and 3 fm 
correspond to two columns. On the left, the partial trigger rate is shown in 
units of 10$^{5}$ s$^{-1}$. On the right, the partial rates are normalized 
to the total trigger rate and shown in units of percent.

\subsection{First level trigger based on M$_{T}$}
\label{ssec:fmt}

The trigger scheme studied so far for the system Au+Au divides the first level 
trigger into two consecutive steps: In a first step, a condition M$_{S}$ on
minimum number of particles in each sector establishes a trigger transition. 
In a second step, the trigger transition is asserted or rejected within a 
subsequent time window depending on whether the total multiplicity condition 
M$_{T}$ is satisfied. However, a one step trigger scheme, based exclusively on 
the total multiplicity M$_{T}$, seems also feasible since the total multiplicity
condition M$_{T}$ is more restrictive than the sector multiplicity condition 
M$_{S}$. Fig. \ref{fig:fig12} displays the trigger time achieved by the one and 
two step trigger. The data shown in Fig. \ref{fig:fig12} represent the timing 
of all events which generate a trigger, irrespective of impact parameter. This 
trigger time is shown for the LOR and corrected MEAN timing in the upper and 
lower part, respectively. The solid line shows the timing in the two step scheme
with conditions M$_{S} \geq 8$ and M$_{T} \geq 110$ as explained above. These 
solid line data points correspond to Fig. \ref{fig:fig8} with the additional 
condition of total multiplicity M$_{T}$. The FWHM of this distribution is about
1 nsec. The dashed line in the lower part of Fig. \ref{fig:fig12} represents the
corrected MEAN timing in the one step trigger. Here, the trigger time is defined
by the timing of the particle for which the total multiplicity condition M$_{T} 
\geq 110$ becomes true. As expected, the timing is delayed compared to the two 
step trigger and considerably broadened. The FWHM of this distribution is about 
6 nsec. The upper part of Fig. \ref{fig:fig12} displays the corresponding data 
in the LOR timing. Similarly as in the corrected MEAN time, the trigger time 
in the one step trigger approach is delayed and broadened.   

\section{Conclusions}

A novel approach for a dielectron trigger improves the first level trigger 
performance considerably. This new approach achieves electron identification by 
using isochronous TOF information. The proposed first level trigger architecture 
combines TOF conditions for the two fastest particles of the event with 
multiplicity requirements. The chosen TOF conditions result in a single electron
efficiency of 99\%. Simulations which take the HADES spectrometer as an 
example show that this new approach reduces the beam time requirement by a 
factor of five in the low multiplicity system p+Ni and by a factor of two 
in the medium multiplicity system Ne+Ne. In the high multiplicity system Au+Au,
a first level trigger based exclusively on multiplicity can be implemented in 
both a one or two step trigger architecture. By judicious choice of 
multiplicity conditions, the number of central collisions passed onto the next 
trigger stage can be maximized. 

\begin{ack}

The support of the electron group at GSI and, in particular, fruitful 
discussions with W.Koenig are gratefully acknowledged. The authors thank 
Gy. Wolf for providing the BUU data files used in the simulations.

\end{ack}

\newpage

\vspace{1.5cm}
{\Large \bf
\noindent
Tables
}

\begin{table}[ht]
\begin{tabular}{||c||c||c|c||c|c||c|c||} \hline
$\sigma_{Start}$&Window&\multicolumn{2}{c||}{T$_{0}$\,=\,0\,$\mu$sec }
&\multicolumn{2}{c||}{T$_{0}$\,=\,6\,$\mu$sec }
&\multicolumn{2}{c||}{T$_{0}$\,=\,10\,$\mu$sec } \\ \cline{3-8}
[nsec]& $\Delta$T$_{0}$[nsec]&rate[10$^5$]&R$_{DT}$&rate[10$^5$]&R$_{DT}$
&rate[10$^5$]&R$_{DT}$\\ \cline{1-8}
    & 60.0 & 4.78 & 1.0 & 1.25 & 0.26 & 0.83 & 0.17 \\ \cline{1-8}
0.0 & 14.1 & 0.19 & 1.0 & 0.17 & 0.90 & 0.16 & 0.84 \\ \hline
0.1 & 14.1 & 0.19 & 1.0 & 0.17 & 0.92 & 0.16 & 0.83 \\ \hline
0.2 & 14.2 & 0.22 & 1.0 & 0.20 & 0.90 & 0.18 & 0.82 \\ \hline
0.4 & 14.4 & 0.28 & 1.0 & 0.24 & 0.85 & 0.22 & 0.78 \\ \hline

\end{tabular}
\vspace{.2cm}
\caption{
Trigger rates from corrected MEAN timing in the system p+Ni for deadtimes 
T$_{0}$ = 0,6 and 10\,$\mu$sec. In the left column, the rates are shown in 
units of 10$^{5}$ s$^{-1}$. In the right column, the deadtime  reduction 
factor R$_{DT}$ is shown. The first row shows the rate without trigger 
time window condition. Subsequent rows show the rates if the correct 
window condition $\Delta$T$_{0}$ is applied. 
\label{tab:tab1}
}
\end{table}

\vspace{.5cm}

\begin{table}[ht]
\begin{tabular}{||c||c||c|c||c|c||c|c||} \hline
$M_{L}$&Window&\multicolumn{2}{c||}{T$_{0}$\,=\,0\,$\mu$sec }
&\multicolumn{2}{c||}{T$_{0}$\,=\,6\,$\mu$sec }
&\multicolumn{2}{c||}{T$_{0}$\,=\,10\,$\mu$sec } \\ \cline{3-8}
& $\Delta$T$_{0}$[nsec]&rate[10$^5$]&R$_{DT}$&rate[10$^5$]&R$_{DT}$
&rate[10$^5$]&R$_{DT}$\\ \cline{1-8}
2 & 60.0 & 3.81 & 1.0 & 1.71 & 0.31 & 0.80 & 0.21 \\ \cline{2-8} 
  & 14.2 & 2.23 & 1.0 & 0.95 & 0.43 & 0.69 & 0.31 \\ \hline
3 & 60.0 & 3.03 & 1.0 & 1.09 & 0.36 & 0.76 & 0.25 \\ \cline{2-8} 
  & 14.2 & 1.36 & 1.0 & 0.75 & 0.55 & 0.58 & 0.43 \\ \hline
4 & 60.0 & 2.50 & 1.0 & 1.01 & 0.40 & 0.72 & 0.29 \\ \cline{2-8} 
  & 14.2 & 0.81 & 1.0 & 0.55 & 0.68 & 0.45 & 0.56 \\ \hline
5 & 60.0 & 2.09 & 1.0 & 0.94 & 0.45 & 0.68 & 0.33 \\ \cline{2-8} 
  & 14.2 & 0.48 & 1.0 & 0.37 & 0.78 & 0.32 & 0.67 \\ \hline
6 & 60.0 & 1.75 & 1.0 & 0.86 & 0.49 & 0.64 & 0.37 \\ \cline{2-8} 
  & 14.2 & 0.26 & 1.0 & 0.23 & 0.86 & 0.21 & 0.80 \\ \hline

\end{tabular}
\vspace{.2cm}
\caption{
Trigger rates from corrected MEAN timing in the system Ne+Ne for deadtimes 
T$_{0}$ = 0,6 and 10\,$\mu$sec and for total multiplicity conditions 
M$_{L}$ = 2,3,4,5 and 6. In the left column, the rates are shown in 
units of 10$^{5}$ s$^{-1}$. In the right column, the deadtime  reduction 
factor R$_{DT}$ is shown. For each multiplicity M$_{L}$, the first and 
second row show the rates without and with trigger time window condition.
\label{tab:tab2}
}
\end{table}

\vspace{.5cm}

\vspace{1.cm}
\begin{table}[ht]
\begin{tabular}{||c||c|c||c|c||c|c||} \hline
Au + Au&\multicolumn{2}{c||}{T$_{0}$\,=\,0\,$\mu$sec } 
&\multicolumn{2}{c||}{T$_{0}$\,=\,6\,$\mu$sec } 
&\multicolumn{2}{c||}{T$_{0}$\,=\,10\,$\mu$sec } \\ \cline{2-7}
&\multicolumn{2}{c||}{rate [10$^{5}$]} 
&\multicolumn{2}{c||}{rate [10$^{5}$]} 
&\multicolumn{2}{c||}{rate [10$^{5}$]} \\ \cline{1-7}
$M_{T} \geq 100$ & 1.08 & 0.37 & 0.90 & 0.31 & 0.80 & 0.28 \\ 
$M_{T} \geq 105$ & 1.08 & 0.28 & 0.94 & 0.24 & 0.85 & 0.22 \\ 
$M_{T} \geq 110$ & 1.08 & 0.19 & 0.98 & 0.17 & 0.91 & 0.16 \\ 
$M_{T} \geq 115$ & 1.08 & 0.13 & 1.01 & 0.12 & 0.97 & 0.12 \\ 
$M_{T} \geq 120$ & 1.09 & 0.08 & 1.04 & 0.08 & 1.01 & 0.07 \\ \hline
\end{tabular}
\vspace{1.cm}
\caption{
Rates of first level trigger transitions (left column) and of triggers 
(right column) in the system Au+Au (see text). The rates are shown in 
units of 10$^{5}$ s$^{-1}$ for deadtimes T$_{0}$ = 0,6 and 10\,$\mu$sec 
and for different total multiplicity conditions M$_{T}$. 
\label{tab:tab3}
}
\end{table}

\vspace{2.cm}
\begin{table}[ht]
\begin{tabular}{||c||c|c||c|c||} \hline
Au + Au&\multicolumn{2}{c||}{rate b=1fm } 
&\multicolumn{2}{c||}{rate b\,=\,3\,fm} \\ \cline{2-5}
& [10$^{5}$] & [\%] & [10$^{5}$] & [\%] \\ \cline{1-5} 
$M_{T} \geq 100$ & 0.11 & 39.7 & 0.17 & 59.8 \\ 
$M_{T} \geq 105$ & 0.11 & 51.0 & 0.11 & 48.6 \\ 
$M_{T} \geq 110$ & 0.11 & 69.5 & 0.05 & 29.9 \\ 
$M_{T} \geq 115$ & 0.09 & 80.3 & 0.02 & 19.0 \\ 
$M_{T} \geq 120$ & 0.07 & 90.3 & 0.01 & 8.7 \\ \hline
\end{tabular}
\vspace{1.cm}
\caption{
First level partial trigger rates of events with impact parameters b\,=\,1 and 
3\,fm in the system Au+Au for different total multiplicity conditions M$_{T}$.
The left column displays the partial rate in units of 10$^{5}$ s$^{-1}$. In 
the right column, the partial rates are normalized to the total trigger rate 
and shown in units of percent.
\label{tab:tab4}
}
\end{table}

\newpage
{\Large \bf
\noindent
Figure Captions
}

\begin{figure}[ht]
\vspace{1.3cm}
\caption{                 
LOR trigger time from the two fastest particle of the event. The solid line 
shows the data of dielectron events. The dashed line displays the data for p+Ni 
events which do not contain dielectrons. The vertical dotted line represents 
the dielectron TOF condition.
}
\label{fig:fig1}
\end{figure}

\begin{figure}[ht]
\vspace{1.3cm}
\caption{                 
TOF data for single electron trajectories as a function of paddle number. The 
diamond symbols represent the TOF values alone. The solid and dashed line data 
represent the time in the LOR and MEAN timing approach, respectively.
}
\label{fig:fig2}
\end{figure}

\begin{figure}[ht]
\vspace{1.3cm}
\caption{               
Corrected LOR and MEAN trigger time from the two fastest particle of the event
are shown in the upper and lower part, respectively. The solid and dashed line
data represent dielectron events and no-pair p+Ni events, respectively.
The vertical lines indicate the dielectron TOF conditions (see text).
}
\label{fig:fig3}
\end{figure}

\begin{figure}[ht]
\vspace{1.3cm}
\caption{                 
The trigger time from the two fastest particle of Ne+Ne events
for impact parameters b\,=\,1,3 and 5\,fm. The corrected LOR and MEAN 
timing are shown in the upper and lower part, respectively. The vertical 
dotted lines indicate the trigger window condition $\Delta$T$_{0}$.
}
\label{fig:fig4}
\end{figure}

\begin{figure}[ht]
\vspace{1.3cm}
\caption{                 
Trigger efficiency for the system Ne+Ne at 2 AGeV as a function of the 
minimum total multiplicity M$_{L}$. Shown are data points for events with 
impact parameters of b=1,2 and 3\,fm.
}
\label{fig:fig5}
\end{figure}

\begin{figure}[ht]
\vspace{1.3cm}
\caption{                 
The trigger timing from the two fastest particle of Au+Au events for impact 
parameters b\,=\,1,3 and 5\,fm. The corrected LOR and MEAN timing are shown 
in the upper and lower part, respectively. The vertical 
dotted lines indicate the trigger window condition $\Delta$T$_{0}$.
}
\label{fig:fig6}
\end{figure}

\begin{figure}[ht]
\vspace{1.3cm}
\caption{                 
First level trigger efficiency for the system Au+Au at 1 AGeV
as a function of the minimum sector multiplicity M$_{S}$.
Shown are data points for events with impact parameters of b=1,3 and 5\,fm.
}
\label{fig:fig7}
\end{figure}

\begin{figure}[ht]
\vspace{1.3cm}
\caption{                 
Corrected MEAN trigger time in the Au+Au system derived from sector timings of 
the two fastest particles in each sector(see text). Shown are the data for 
impact parameters b=1,3 and 5\,fm. The time zero is the time of reaction.
}
\label{fig:fig8}
\end{figure}

\begin{figure}[ht]
\vspace{1.3cm}
\caption{                 
The maximum total Au+Au event multiplicity for impact parameters b=1,3 and 
5\,fm reached during a time window of 20\,nsec following the trigger transition.
}
\label{fig:fig9}
\end{figure}

\begin{figure}[ht]
\vspace{1.3cm}
\caption{
Time at which maximum total event multiplicity is reached for impact parameters
b=1,3 and 5\,fm. The time zero is the time of trigger transition.
}
\label{fig:fig10}
\end{figure}

\begin{figure}[ht]
\vspace{1.3cm}
\caption{                 
Efficiency of first level trigger (deadtime losses included) for central
events. Shown are the data points as a function of the total multiplicity 
condition M$_{T}$ for deadtimes of 0,6 and 10\,$\mu$sec.
}
\label{fig:fig11}
\end{figure}

\begin{figure}[ht]
\vspace{1.3cm}
\caption{                 
Trigger timing in the one (dashed line) and two step (solid line) trigger 
scheme (see text). The data in the LOR and corrected MEAN timing are  shown in 
the upper and lower part, respectively. The time zero is the time of reaction.
}
\label{fig:fig12}
\end{figure}

\clearpage

\vspace{0.cm}

\newpage

\pagestyle{empty}

\begin{minipage}{12.cm}
\epsfig{figure=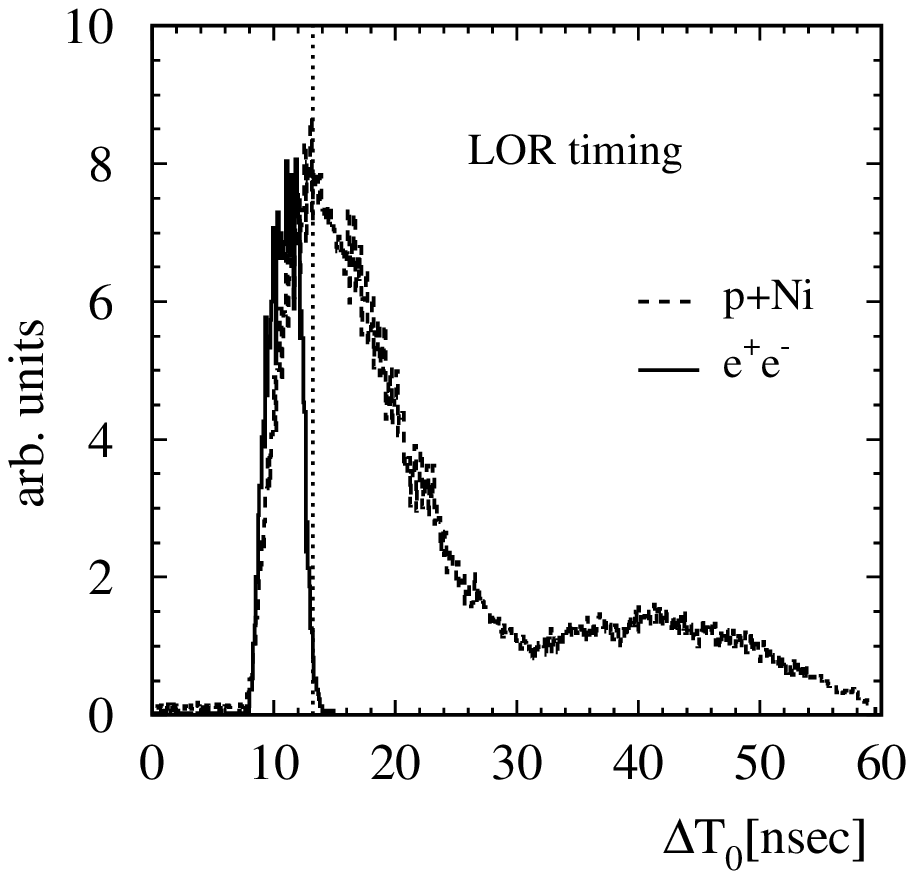,width=.88\linewidth}
\hfill
\end{minipage}

\begin{minipage}{2.cm}
\vspace{0.cm}
{\bf\huge FIG.1}
\end{minipage}

\begin{minipage}{12.cm}
\epsfig{figure=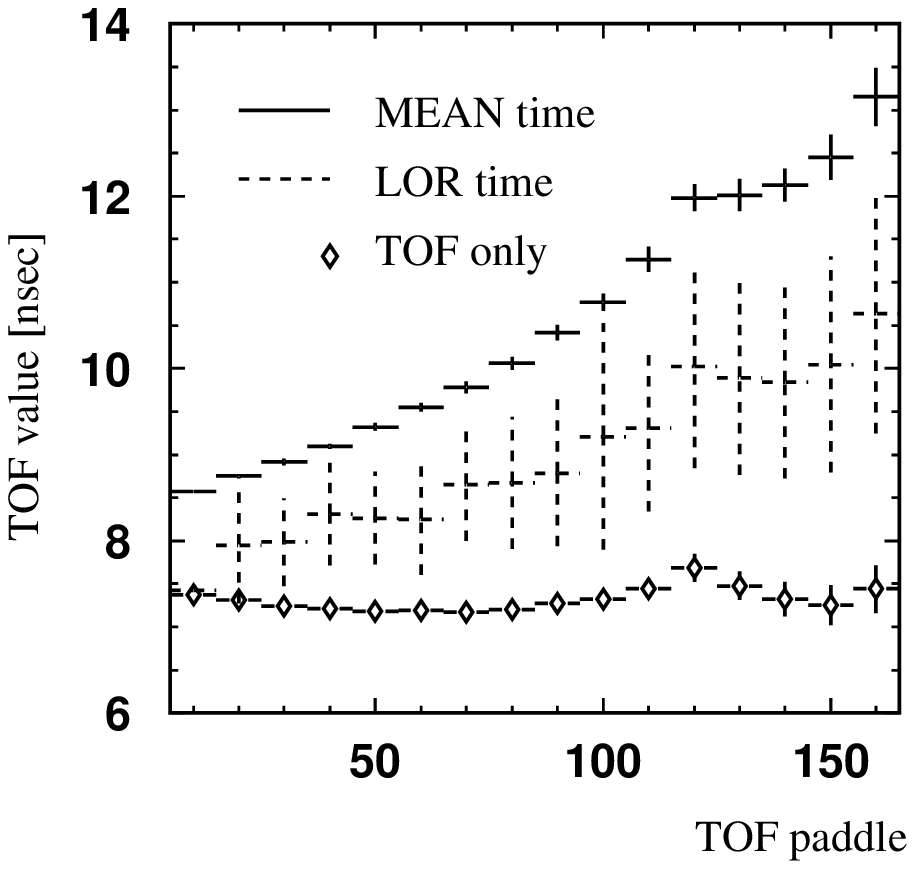,width=.88\linewidth}
\hfill
\end{minipage}

\begin{minipage}{2.cm}
\vspace{0.cm}
{\bf\huge FIG.2}
\end{minipage}

\newpage

\begin{minipage}{12.cm}
\epsfig{figure=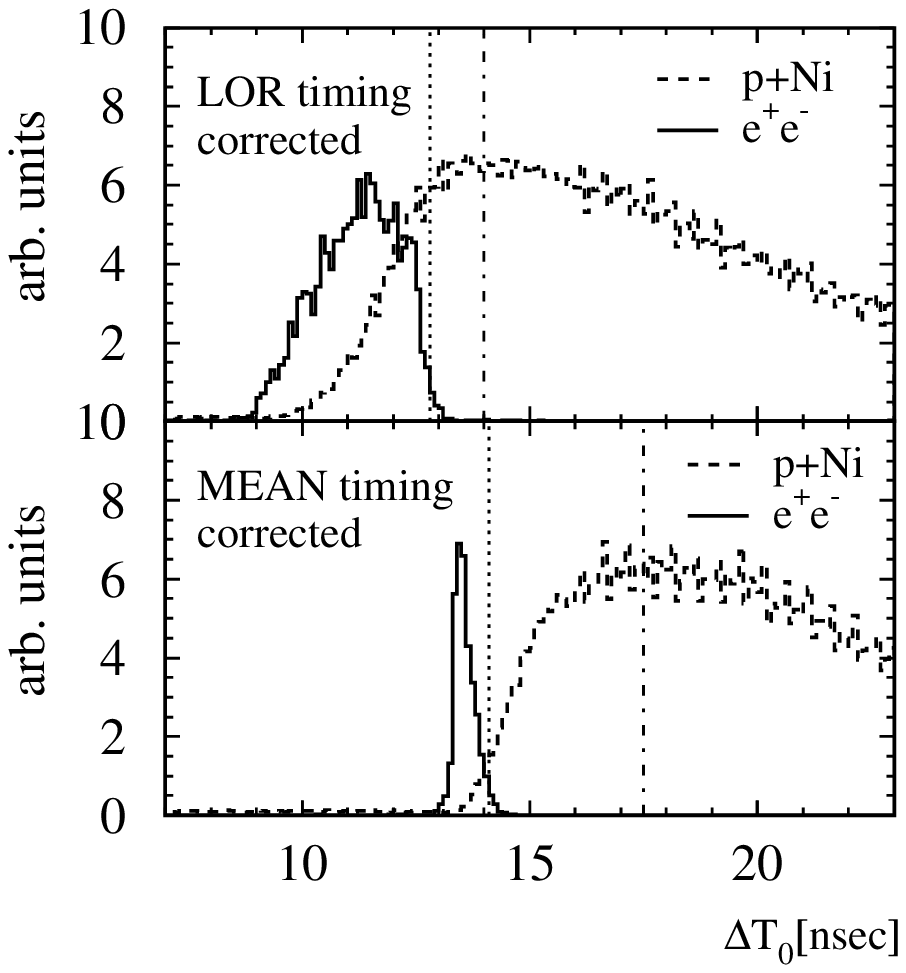,width=.88\linewidth}
\hfill
\end{minipage}

\begin{minipage}{2.cm}
\vspace{0.cm}
{\bf\huge FIG.3}
\end{minipage}

\begin{minipage}{12.cm}
\epsfig{figure=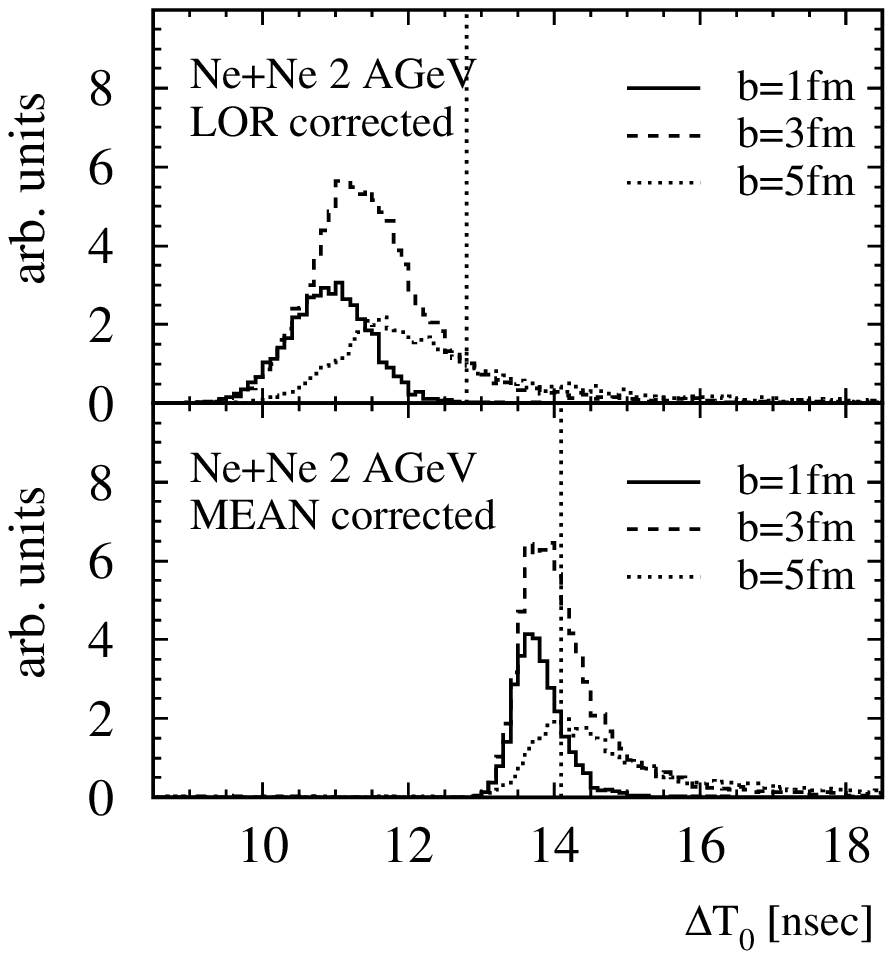,width=.88\linewidth}
\hfill
\end{minipage}

\begin{minipage}{2.cm}
\vspace{0.cm}
{\bf\huge FIG.4}
\end{minipage}

\newpage

\begin{minipage}{12.cm}
\epsfig{figure=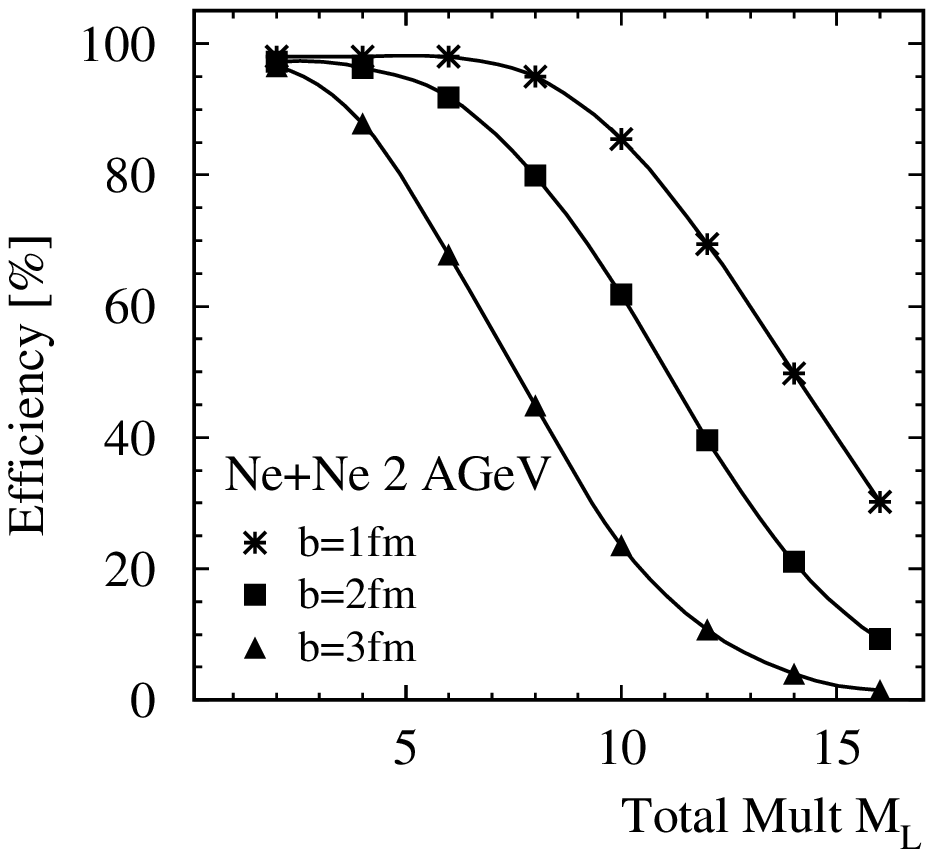,width=.88\linewidth}
\hfill
\end{minipage}

\begin{minipage}{2.cm}
\vspace{0.cm}
{\bf\huge FIG.5}
\end{minipage}

\begin{minipage}{12.cm}
\epsfig{figure=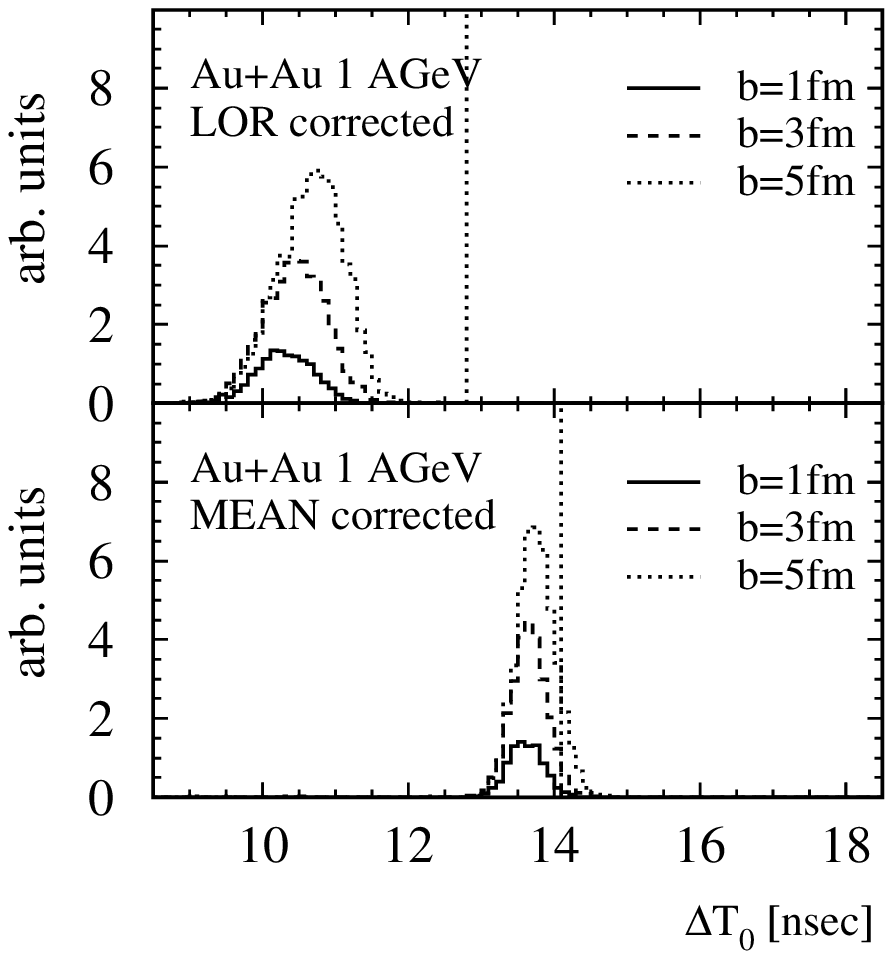,width=.88\linewidth}
\hfill
\end{minipage}                

\begin{minipage}{2.cm}
\vspace{0.cm}
{\bf\huge FIG.6}
\end{minipage}

\newpage

\begin{minipage}{12.cm}
\epsfig{figure=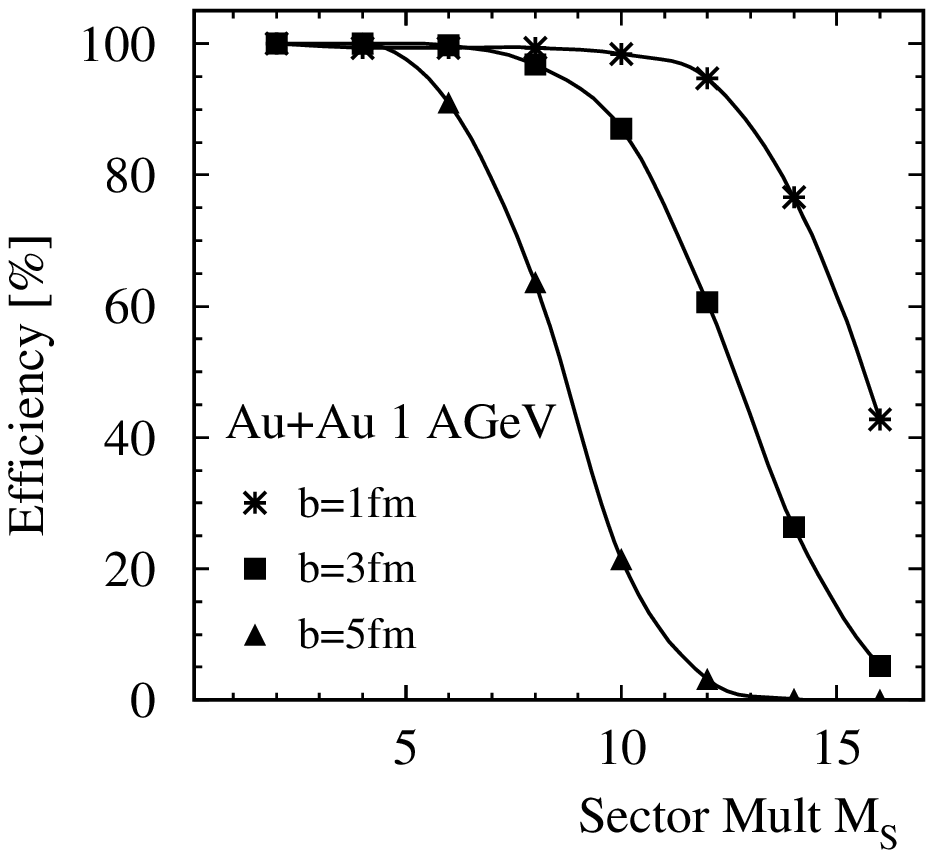,width=.88\linewidth}
\hfill
\end{minipage}

\begin{minipage}{2.cm}
\vspace{0.cm}
{\bf\huge FIG.7}
\end{minipage}

\begin{minipage}{12.cm}
\epsfig{figure=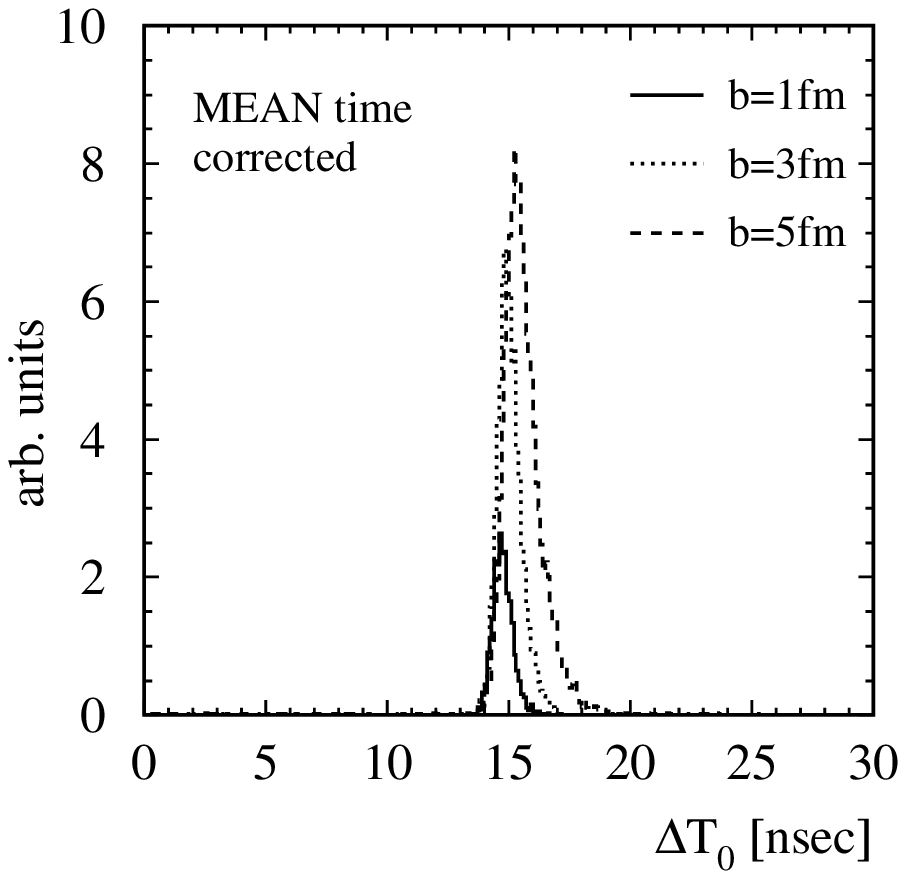,width=.88\linewidth}
\hfill
\end{minipage}

\begin{minipage}{2.cm}
\vspace{0.cm}
{\bf\huge FIG.8}
\end{minipage}

\newpage

\begin{minipage}{12.cm}
\epsfig{figure=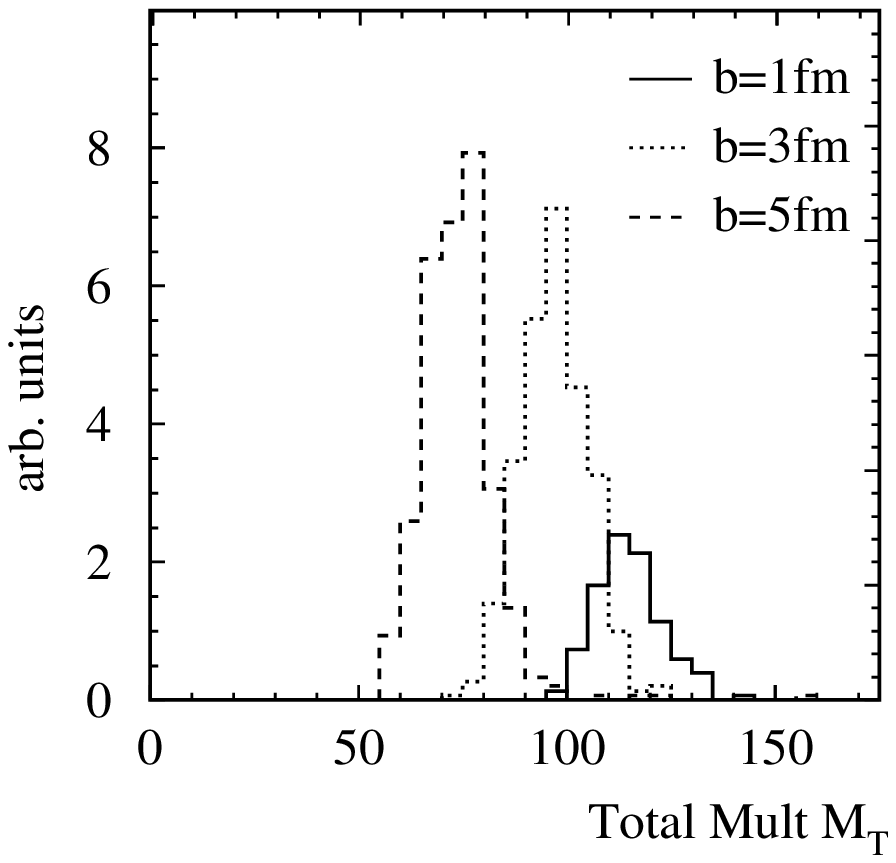,width=.88\linewidth}
\hfill
\end{minipage}

\begin{minipage}{2.cm}
\vspace{0.cm}
{\bf\huge FIG.9}
\end{minipage}

\begin{minipage}{12.cm}
\epsfig{figure=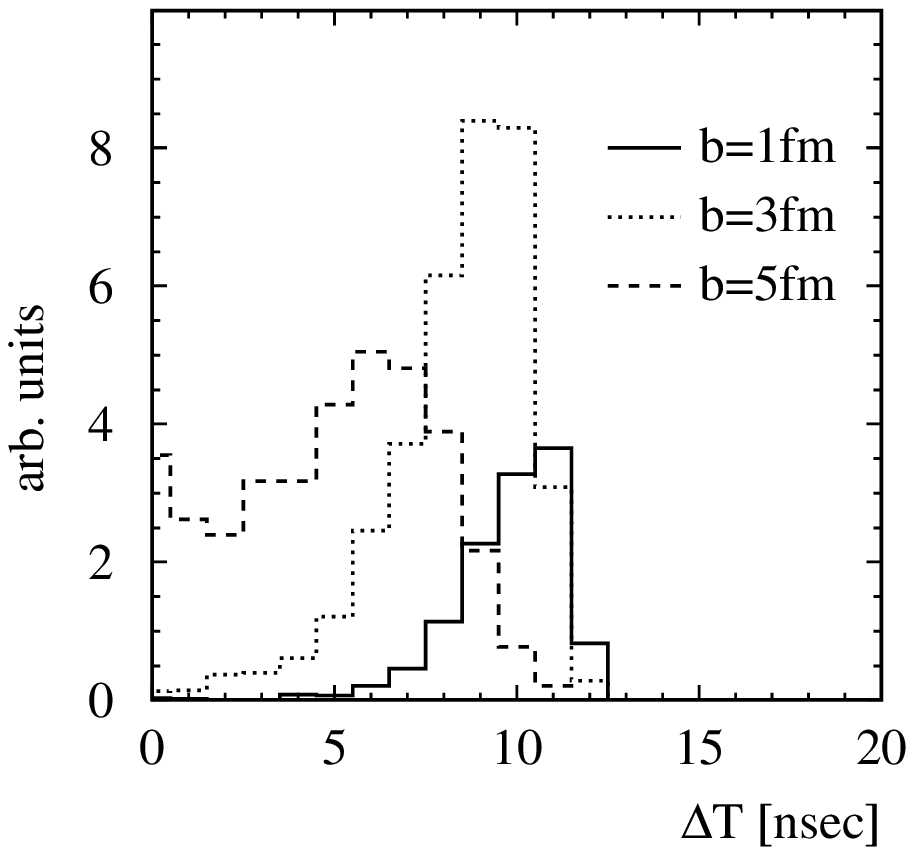,width=.88\linewidth}
\hfill
\end{minipage}

\begin{minipage}{2.cm}
\vspace{0.cm}
{\bf\huge FIG.10}
\end{minipage}

\newpage

\begin{minipage}{12.cm}
\epsfig{figure=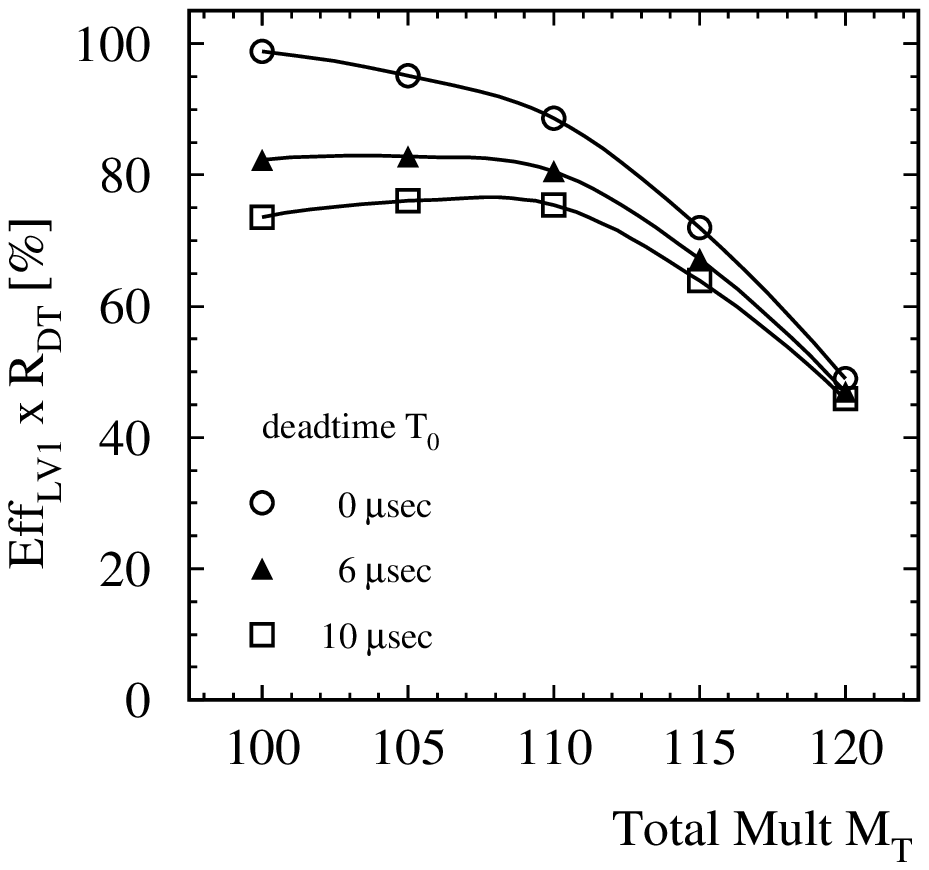,width=.88\linewidth}
\hfill
\end{minipage}

\begin{minipage}{2.cm}
\vspace{0.cm}
{\bf\huge FIG.11}
\end{minipage}

\begin{minipage}{12.cm}
\epsfig{figure=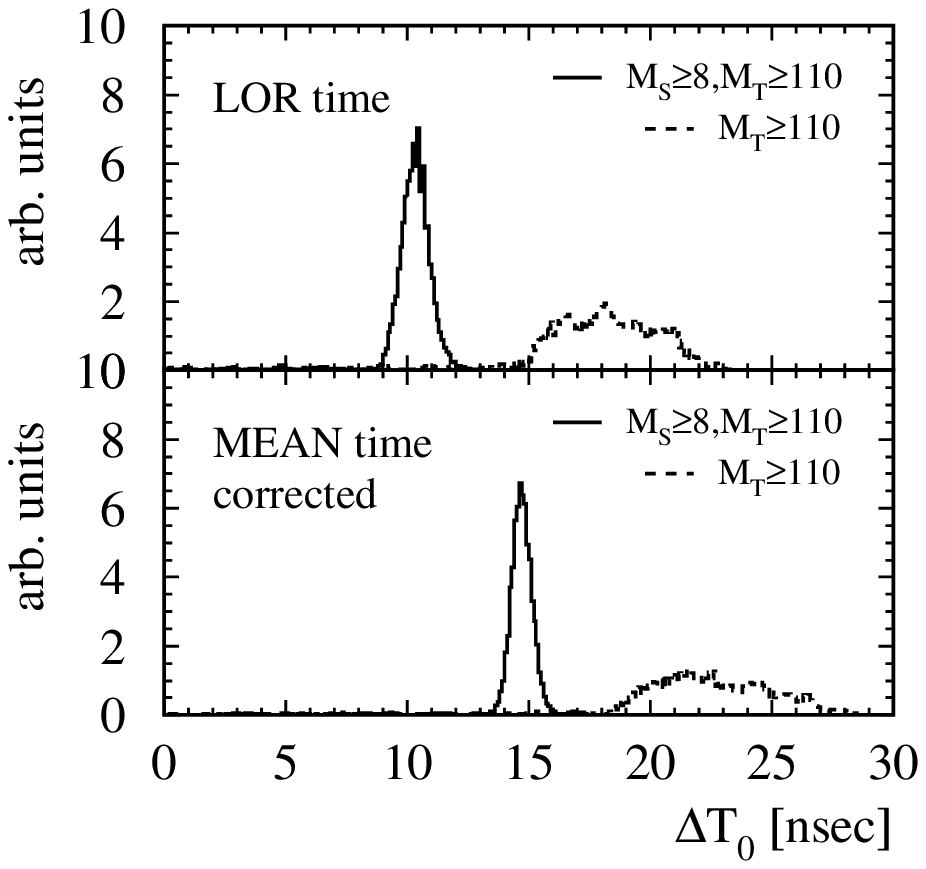,width=.88\linewidth}
\hfill
\end{minipage}

\begin{minipage}{2.cm}
\vspace{0.cm}
{\bf\huge FIG.12}
\end{minipage}

\end{document}